\documentclass[aps,prb,twocolumn,showpacs,superscriptaddress,floatfix,nofootinbib]{revtex4}
\usepackage{amsfonts,hyperref}

\usepackage{graphicx}
\usepackage{amsmath}
\usepackage{amssymb}

\newcommand{\hc}{\hat c}

\begin{document}
\title{Stripes in the two-dimensional $t$-$J$ model with infinite projected entangled-pair states}
\author{Philippe Corboz}
\affiliation{Theoretische Physik, ETH Zurich, 8093 Zurich, Switzerland} 
\affiliation{Institut de th\'eorie des ph\'enom\`enes physiques, \'Ecole Polytechnique F\'ed\'erale de Lausanne, CH-1015 Lausanne, Switzerland} 

\author{Steven R. White}
\affiliation{Department of Physics and Astronomy, University of California, Irvine, CA 92697-4575 USA}

\author{Guifr\'e Vidal} 
\affiliation{School of Mathematics and Physics, The University of Queensland, QLD 4072, Australia}

\author{Matthias Troyer}
\affiliation{Theoretische Physik, ETH Zurich, 8093 Zurich, Switzerland}

\date{\today}

\begin{abstract}
We simulate the $t$-$J$ model in two dimensions by means of infinite projected entangled-pair states (iPEPS) generalized to arbitrary unit cells, finding results similar to those previously obtained by the density-matrix renormalization group (DMRG) for wide ladders.  In particular, we show that states exhibiting stripes, {\it i.e}. a unidirectional modulation of hole-density and antiferromagnetic order with a $\pi$-phase shift between adjacent stripes, have a lower variational energy than uniform phases predicted by variational and fixed-node Monte Carlo simulations. For a fixed unit-cell size the energy per hole is minimized for a hole density $\rho_l\sim 0.5$ per unit length of a stripe. The superconducting order parameter is maximal around $\rho_l\sim 0.75-0.8$.   
\end{abstract}


\maketitle
\section{Introduction}
The simulation of strongly correlated electron systems in two dimensions remains one of the biggest challenges in condensed matter physics. The infamous negative sign problem \cite{sign} prevents accurate quantum Monte Carlo simulations of large fermionic systems at low temperature. An alternative route without a sign problem is provided by variational Monte Carlo (VMC). 
Typically, several ansatz wave functions, biased towards different orders, are optimized and the one with lowest energy is considered as the best approximation to the ground state. A powerful way to improve upon a variational wave function is to use it as a guiding wave function for the fixed-node Monte Carlo (FNMC) method, \cite{FNMC,Lugas06} which yields the best variational wave function with the same nodal structure as the guiding wave function.

In recent years, a new class of variational wave functions for two-dimensional systems have been proposed: the so-called tensor network states, including e.g. the multi-scale entanglement renormalization ansatz (MERA) \cite{MERA} and projected entangled-pair states (PEPS), \cite{PEPS} also called tensor product states (TPS).\cite{TPS} These tensor networks have recently been generalized to fermionic systems,\cite{FTN} and can be seen as generalizations of the density-matrix renormalization group (DMRG)\cite{White92} method to two dimensions. 
In contrast to other variational wave functions, tensor network states are largely unbiased,\cite{Corboz10b} 
with an accuracy which can be systematically controlled by the so-called bond dimension $D$ (typically called $m$ in DMRG). 
DMRG yields very accurate results for  quasi one-dimensional ladder systems with cylindrical boundary conditions up to a width around $8-12$, but becomes numerically inefficient for larger widths. 
The computational cost of MERA and PEPS is polynomial in system size and in $D$, however, with such a high power in $D$ that one may question if these methods are competitive to solve hard problems in condensed matter physics.

In this paper we show that tensor networks indeed can compete with the best known variational methods, in particular with FNMC based on Gutzwiller projected ansatz wave functions.\cite{Lugas06} Specifically, we simulate the doped $t$-$J$ model for $J/t=0.4$ in the thermodynamic limit  
with infinite PEPS (iPEPS \cite{iPEPS}) and with DMRG for systems up to a width $10$, 
and find significantly lower variational energies than obtained with FNMC. It was previously shown that iPEPS yields lower energies than VMC,\cite{Corboz10b,Corboz10c} but the values were still higher than  the ones from FNMC. 
Here we find that extending the ansatz to larger unit cells than the $2\times2$ cells previously used leads to a considerable improvement of the variational energy, which is an indication that the ground state may break translational invariance on a larger scale than $2\times2$.\cite{commentSB, commentHuan}  
By inspection of local order parameters we find that the ground state of the two-dimensional $t$-$J$ model exhibits stripes, {\it i.e.} a unidirectional modulation of the hole density and the antiferromagnetic order, as previously found with DMRG in cylinders up to width 8.\cite{DMRGstripes, White98b} This is in contrast to the findings from VMC \cite{VMCstudies, Lugas06} and FNMC,\cite{Lugas06} which have suggested a uniform phase.  

\section{Method}
\begin{figure}[]
\begin{center}
\includegraphics[width=8.5cm]{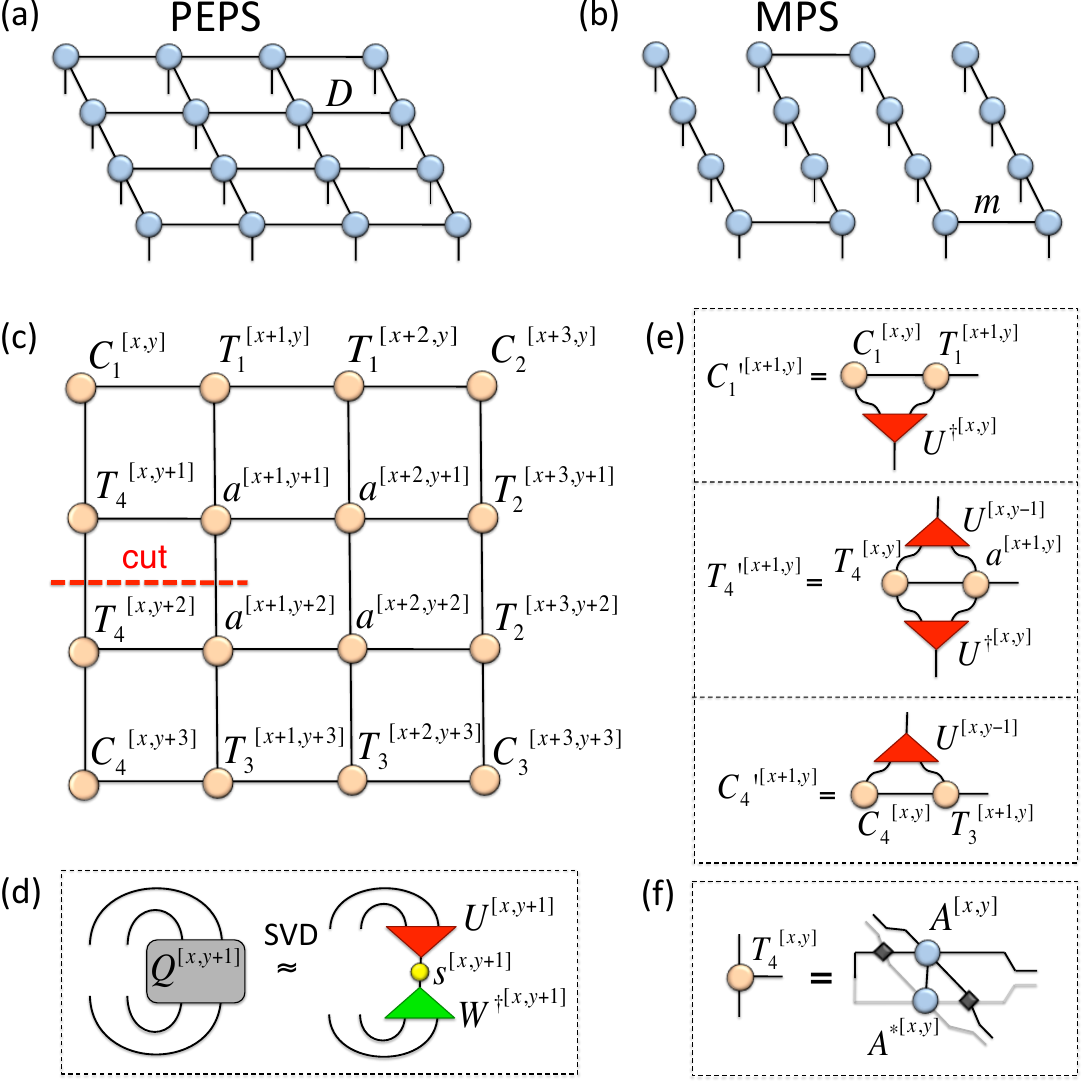}
\caption{(Color online) a) Diagrammatic representation of a PEPS with bond dimension $D$ on a $4\times 4$ lattice. b) An MPS with bond dimension $m$ to represent a $4\times 4$ system. 
c)-e) Diagrams for the left move of the corner-transfer matrix method: (c) A plaquette of four reduced tensors $a$ embedded in the environment. The coordinates $[x,y]$ are to be understood relatively to the unit cell, i.e. one has to take the coordinate $x$ ($y$) modulo $N_x$ (modulo $N_y$). Cutting through the lines as marked in the figure yields the tensor $Q^{[x,y+1]}$ in panel (d): a singular value decomposition of tensor $Q^{[x,y+1]}$ is performed, where only the $\chi$ largest singular values are kept. The resulting isometry $U^{[x,y+1]}$ and its conjugate are used as an approximate resolution of the identity, $\mathrm{I} \approx U^{\dagger [x,y+1]} U^{[x,y+1]}$. For a fixed $x=x_0$ one computes all isometries $U^{[x_0,y]}$ for all $y \in [1,N_y]$. These isometries are then used to obtain the renormalized corner tensors $C_1'^{[x+1,y]}$ and $C_4'^{[x+1,y]}$, and edge tensors $T_4'^{[x+1,y]}$ for all $y \in [1,N_y]$ and fixed $x=x_0$, shown in e). This whole procedure is repeated $N_x$ times for $x_0 \in [1,N_x]$ to complete an entire left move. (f) Initialization of a boundary tensor from a PEPS tensor and its conjugate, where crossings have been replaced by swap tensors (cf. Ref.~\onlinecite{Corboz10b}). }
\label{fig:ctm}
\end{center}
\end{figure}

Fermionic iPEPS has been introduced and explained in detail in Ref.~\onlinecite{Corboz10b} and we here repeat only the basic ideas.
A PEPS, illustrated in Fig.~\ref{fig:ctm}a), can be seen as an extension of a matrix-product state (MPS) in Fig.~\ref{fig:ctm}b), the tensor network DMRG is based on, to two dimensions. Each blue-filled circle represents a tensor with a rank given by the number of legs attached to it, where the open leg corresponds to a physical index carrying the local Hilbert space of a lattice site, and the connecting lines are bond indices with a certain bond dimension $m$ or $D$, which characterizes the number of variational parameters in the ansatz. Tracing over all bond indices yields the coefficients of the state 
in the tensor product basis of the local Hilbert spaces of all sites. MPSs and PEPSs enable an efficient representation of states obeying the area law of the entanglement entropy \cite{Eisert09} in one and two dimensions, respectively. 
An MPS can also be used to represent states in two dimensions, e.g. by using a snake structure as in Fig.~\ref{fig:ctm}b), however, the required bond dimension $m$ of the tensors grows exponentially with the system's width, whereas in a PEPS the required $D$ is independent of system size (in the limit of large systems). 
In the present work we consider a PEPS of infinite size (iPEPS) made of a periodically repeated rectangular unit cell containing $L_x \times L_y = N$ different tensors, $A^{[x,y]}$, labeled by the coordinates relative to the unit cell. To obtain an approximate representation of the ground state we perform an imaginary time evolution of an initial, randomly chosen, iPEPS. To efficiently compute observables an approximate contraction scheme, discussed below, is used to evaluate the trace over all bond indices, where the accuracy can be controlled by another parameter called boundary dimension~$\chi$.

%
For the experts, we briefly outline how to treat large unit cells, where we adopt the notation from Refs.~\onlinecite{Corboz10b,Corboz10c}. First, for the imaginary time evolution we use the simple update\cite{Jiang08,Corboz10b} on all bonds in the unit cell. Second, to contract the iPEPS we use the corner-transfer matrix (CTM) method \cite{BaxterCTM,Nishino96}, generalized to larger unit cells. The CTM method yields the so-called environment, consisting of  corner tensors $C_1, C_2, C_3, C_4$ and edge tensors $T_1, T_2, T_3, T_4$,  which account for the infinite system surrounding the tensors in the ``bulk" of the system (Fig.~\ref{fig:ctm}c)). 
We again assign coordinates $[x,y]$ to each of these tensors to label the relative position in the unit cell, i.e. $4N$ corner tensors and $4N$ edge tensors in total are separately stored.
Initially, the corner and edge tensors at position $[x,y]$ are constructed similarly as a reduced tensor $a^{[x,y]}$, by multiplying the tensor $A^{[x,y]}$ to its conjugate and fusing the bond indices,\cite{Corboz10b} where we trace over the legs directed toward an open boundary, as illustrated in Fig.~\ref{fig:ctm}f).
The environment tensors are iteratively built by four directional coarse-graining moves (left, right, top, bottom), similarly as proposed in Ref.~\onlinecite{Orus09}. An entire CTM step consists of $L_x$ left moves, $L_x$ right moves, $L_y$ top moves and finally $L_y$ bottom moves. This sequence is repeated until convergence is reached. The renormalization procedure, adopted from Refs.~\onlinecite{Nishino96, Corboz10c}, is based on a $4\times4$ cell of tensors to compute an appropriate isometry. These isometries are used to absorb a column (or row) of tensors into the corresponding boundary tensors, which effectively corresponds to the growth of the system by one column (or row) of sites.  
The left move is explained in Fig.~\ref{fig:ctm}, and one proceeds similarly for the other moves. 

\section{Simulation results}
With the generalized iPEPS introduced in the last section we simulate the $t$-$J$ model, given by the Hamiltonian
\begin{equation}
H= - t \sum_{\langle ij \rangle \sigma} \left( \tilde{c}_{i \sigma}^{\dagger}\tilde{c}_{j\sigma}  + H.c.\right) +  J\sum_{\langle ij \rangle}  \left( \hat S_i \hat S_j - \frac{1}{4} \hat n_i \hat n_j\right)  
\end{equation}
with $\sigma=\{\uparrow,\downarrow\}$ the spin index, $\hat n_i=\sum_\sigma \hc^\dagger_{i \sigma} \hc_{i \sigma}$ the electron density and $\hat S_i$ the spin $1/2$ operator on site $i$, and $\tilde{c}_{i\sigma}=\hc_{i\sigma} ( 1 - \hc^\dagger_{i \bar \sigma} \hc_{i \bar \sigma})$, where we fix $J/t=0.4$. 

\begin{figure}[]
\begin{center}
\includegraphics[width=8cm]{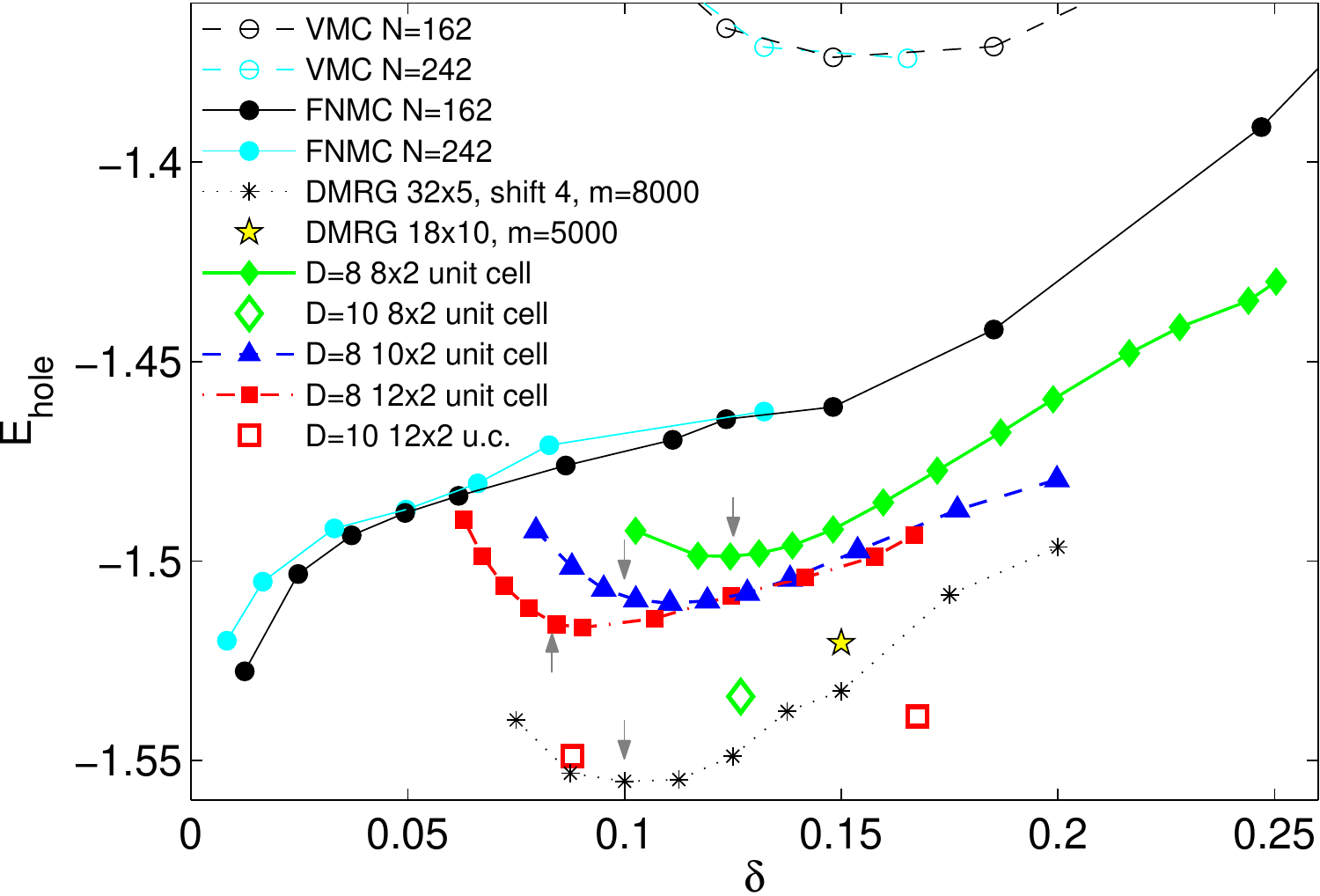}
\caption{(Color online) Energy per hole as a function of doping for $J/t=0.4$. For the $D=8$ results the estimated error due to a finite $\chi=100$ is $0.3\%$. For $D=10$ the values are extrapolated in $\chi$. Grey arrows mark stripes with 0.5 holes per unit length. The DMRG cylinders labeled with a shift  are wrapped periodically with the indicated shift to connect the transverse stripes into one long continuous spiral stripe, allowing arbitrary filling.
}
\label{fig:Ehole}
\end{center}
\end{figure}

Our main result for the infinite two-dimensional $t$-$J$ model is that, in contrast to the uniform phase found by VMC and FNMC,\cite{Lugas06} we find a striped phase in the doping regime $1/12 < \delta <1/4$, in agreement with DMRG calculations on wide ladders.\cite{DMRGstripes} Figure~\ref{fig:Ehole} shows that the variational energies obtained with iPEPS and DMRG are considerably lower than the ones from VMC and FNMC. We consider here  the energy per hole (in units of $t$), $E_{hole}=(E_s - E_0)/\delta$, with $E_s$ the energy per site and $E_0=-0.467775$ the value at zero doping taken from Ref.~\onlinecite{Sandvik97}. The VMC and FNMC energies are seen to increase slightly with system size, and thus, we expect that in the thermodynamic limit the striped states are energetically favored. 

%
\begin{figure}[]
\begin{center}
\includegraphics[width=8.5cm]{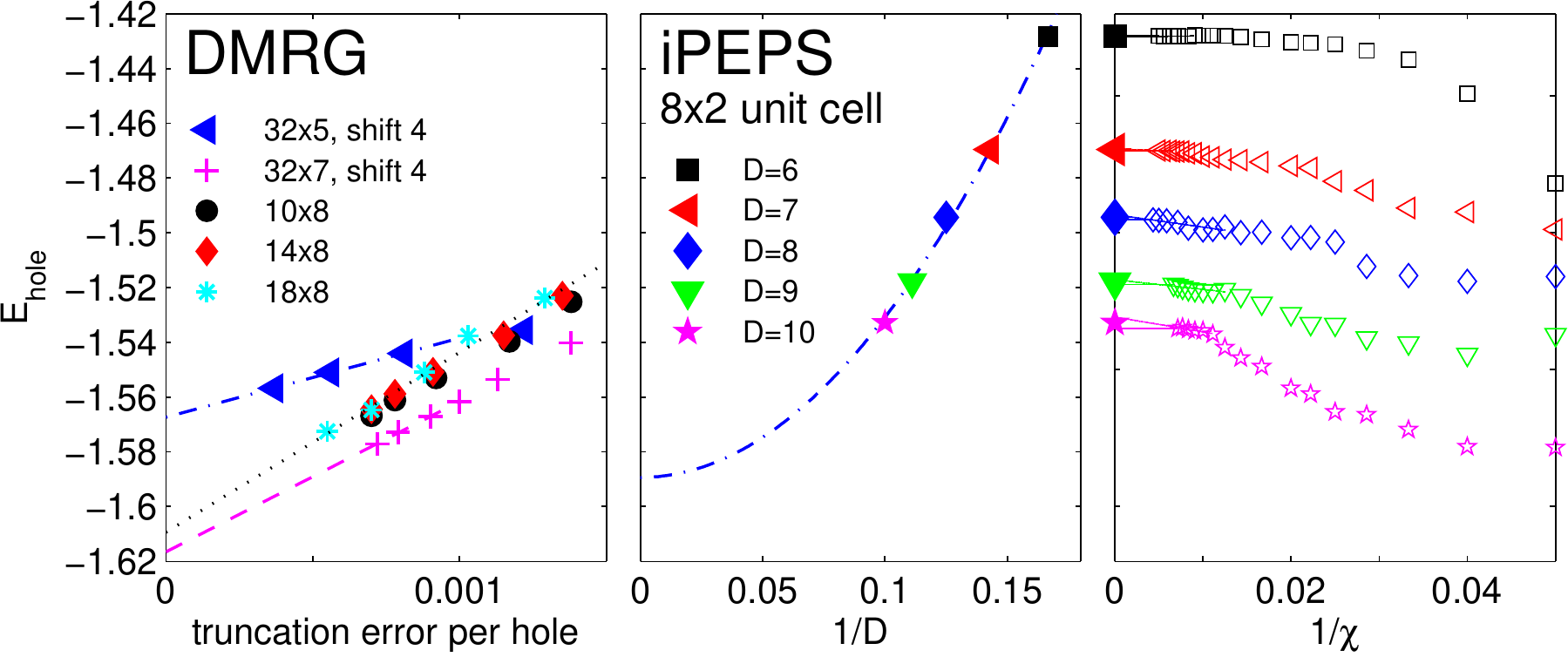}
\caption{(Color online) Energy per hole at doping $\delta \approx 1/8$ for $J/t=0.4$ as a function of truncation error in DMRG (left panel) and as a function of $1/D$ with iPEPS using a $8\times2$ unit cell (middle panel), with values obtained from extrapolation in the  dimension $\chi$ (right panel). The quadratic fit in the middle panel is a guide to the eye. 
}
\label{fig:Echi}
\end{center}
\end{figure}
In  Fig.~\ref{fig:Echi} we present a convergence study  of the iPEPS energies as a function of the dimensions $D$ and $\chi$, and compare it to DMRG results for ladders for a doping $\delta\sim 1/8$.
While iPEPS is a variational ansatz the resulting energies may be non-variational, because of  the approximate contraction of the iPEPS, which introduces an error depending on $\chi$. 
The convergence study in the right panel in Fig. \ref{fig:Echi} indicates that 
this error is smaller than the symbol size, {\it i.e.} the upper end of the symbol provides an upper bound of the ground state energy.
The middle panel shows the values, extrapolated in $\chi$, as a function of $1/D$. The values have not yet converged, thus the ansatz can still be further improved by increasing $D$. Attempting a quadratic fit yields $E_{hole} = - 1.59(3)$ as $D\rightarrow\infty$, which is similar to the extrapolated DMRG results, $E_{hole} = - 1.61(2)$.

\begin{figure}[]
\begin{center}
\includegraphics[width=5.4cm]{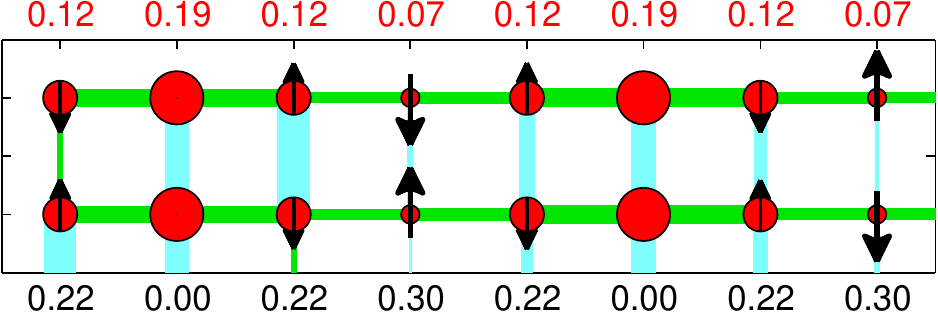} \vspace{0.1cm}

\includegraphics[width=8cm]{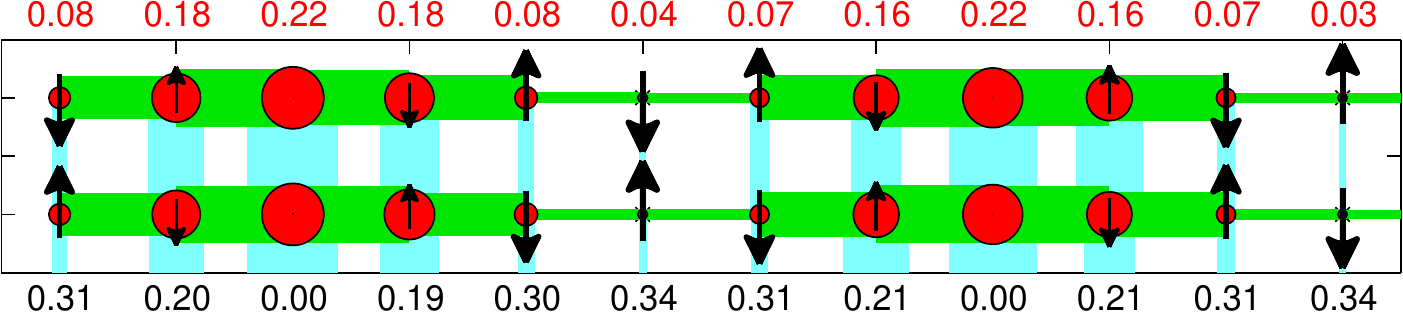} \vspace{0.1cm}

\includegraphics[width=8cm]{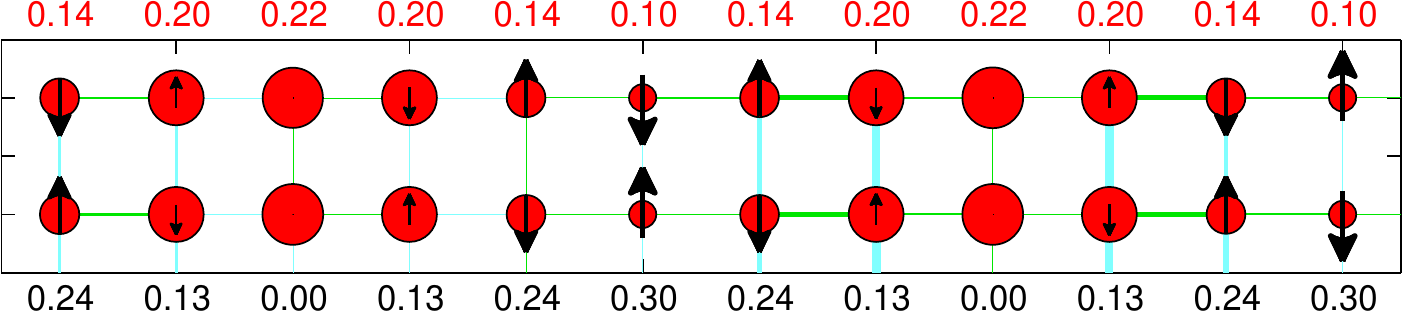}
\caption{(Color online) Examples of stripes running in the vertical direction obtained from various iPEPS simulations. Each panel shows one unit cell of the infinite lattice. The diameter of the dots scales with the local hole density with average values given by the (upper) red numbers. The arrows represent the local magnetic moment with average magnitude given by the (lower) black numbers. There is a $\pi$-phase shift in the antiferromagnetic order between adjacent stripes. The width of a bond between two sites scales with the (singlet) pairing amplitude on the bond with a positive (green) or negative (cyan) sign.  A pattern with predominantly $d$-wave order is visible, with maximal pairing amplitude 0.01, 0.03, and 0.003 in the three examples, respectively.
a) Stripes of width 4 in a unit cell $8\times2$ with hole density of $\rho_l\sim 0.5$ holes per unit length per stripe at a doping $\delta\sim1/8$. 
b) Stripes of width 6 in a unit cell $12\times2$ with $\rho_l\sim 0.75$, $\delta \sim 1/8$, where the pairing is maximal. 
c) Same as in b) but with $\rho_l=1$, $\delta\sim1/6$ where the pairing is suppressed. 
}
\label{fig:examples}
\end{center}
\end{figure}

Examples of stripes obtained with iPEPS are  presented in Fig.~\ref{fig:examples}. These stripes appear without biasing the initial iPEPS to stripe order (we typically start from several random initial states). We tested different unit cell sizes with $L_y \in\{2,4\}$ and $L_x \in \{2,4,5,6,7,8,10,12\}$, where each unit cell imposes a different periodicity on the wave function. Some of the $L_y=4$ samples exhibit a slight modulation also in $y$ direction, however, since the energy is similar to the $L_y=2$ samples we focus on the latter. 

Figure~\ref{fig:Ehole} shows that, for a fixed unit cell size, the energy minimum is found near $\rho_l\sim0.5$ holes per unit length per stripe, e.g. at $\delta=1/8$ for the $8\times 2$ unit cell, which is in agreement with DMRG results (black stars and Ref.~\onlinecite{White98b}). 
However, the minimum for the $8\times 2$ cell at $\delta \sim 1/8$ is higher than the energy of the $10\times2$ or the $12\times2$ sample at the same doping, which indicates that the repulsion between the stripes of width 4 in the $8\times2$ cell is too strong so that stripes with larger widths and larger $\rho_l$ are energetically favored, in contrast to the predictions by DMRG.\cite{White98b} For densities $0.75<\rho_l<1$ we observe the tendency that mixed stripes (two stripes with different densities) yield a lower variational energy than two stripes of equal density, however, we do not observe signs of phase separation between stripes with $\rho_l=1$ and $\rho_l=0.5$ as found with DMRG \cite{White98b}. A future study with larger unit cells and larger $D$ will shed further light in this issue.

An open question is what happens at smaller doping than $1/12$.  
The minimum at $\rho_l$ for different unit cells is seen to decrease with increasing unit cell length $L_x$. It is conceivable that this trend continues, so that with decreasing doping we obtain stripes which are increasingly more widely spaced,\cite{White98b} with a distance between the stripes varying as $d\approx 1/(2\delta)$.\cite{White98b} However, from the present data  we cannot rule out other phases, such as a uniform phase at small doping $\delta<1/12$, or phase separation between an undoped and a doped region. 

\begin{figure}[]
\begin{center}
\includegraphics[width=8cm]{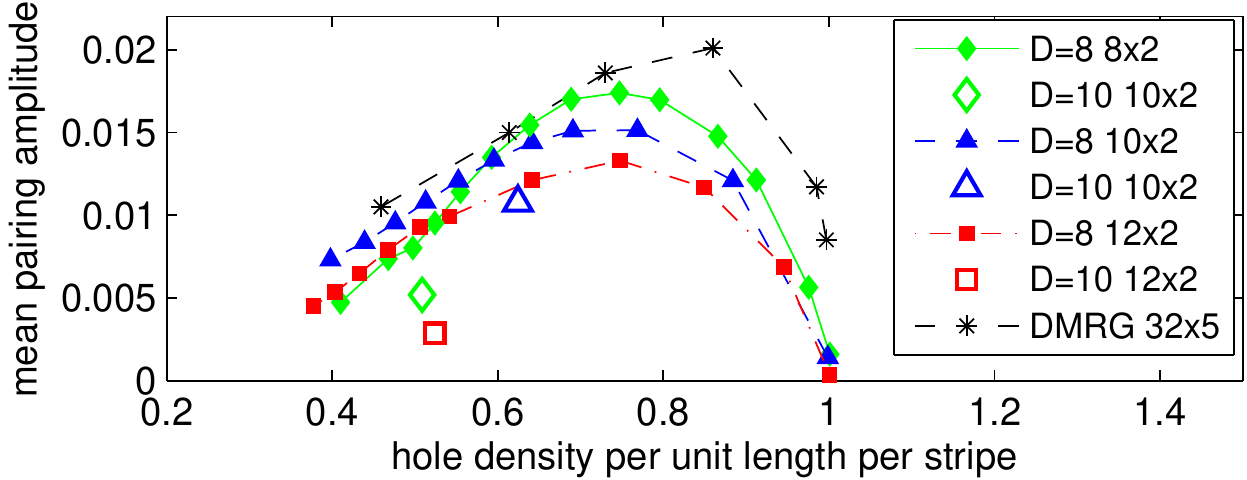}
\caption{(Color online) Mean pairing amplitude as a function of linear hole density per unit length of a stripe $\rho_l$.   The DMRG results are for one longitudinal stripe with a pairing field of 0.02 applied throughout, making the overall magnitude somewhat arbitrary.
}
\label{fig:pairing}
\end{center}
\end{figure}

Finally, we study the pair field  $\Delta=1/\sqrt{2}\langle \hat c_i \hat c_j -\hat c_j\hat c_i \rangle$ between nearest-neighbor sites $(i,j)$ which is modulated along the $x$-direction and predominantly forms a $d$-wave pattern as shown in Fig.~\ref{fig:examples}. The $d$-wave order between neighboring stripes has the same phase, however, we have also observed states with a similar energy where the $d$-wave order exhibits a $\pi$-phase shift between neighboring stripes. Thus, it seems that this $\pi$-phase shift has only little influence on the energy, and from the present data we cannot determine which state is preferred.
The mean pairing amplitude (averaged over the unit cell) in Fig.~\ref{fig:pairing} exhibits a maximum around $\rho_l = 0.75$ ($\rho_l=0.8$ in DMRG).
For $\rho_l=1$ the stripes are insulating with a vanishing pairing amplitude. For $\rho_l=0.5$ the pairing amplitude for $D=8$ is finite, but decreases quickly with increasing $D$, and is possibly insulating for larger $D$, too. In between these two insulating states, excess holes (or electrons) form pairs, leading to an increase of the pairing.
%

\section{Conclusion}
In this paper we presented iPEPS simulation results for the $t$-$J$ model in the thermodynamic limit, where we obtained striped states, in qualitative agreement with DMRG calculations on wide ladders.    
The iPEPS variational energies are considerably lower than the ones from the uniform states    
obtained with VMC and FNMC based on Gutzwiller projected ansatz wave functions, and compatible with DMRG results for finite systems. This demonstrates that iPEPS is a competitive variational method for strongly correlated electron systems. \cite{commentMERA}
The differences between iPEPS and DMRG only involve
quantitative details, such as the precise stripe linear filling and spacing as a function of overall filling. 
 Using larger unit cells  in iPEPS than the usual $2\times2$ is essential to study ground states which break translational invariance on a larger scale than only 2 lattice sites.

The iPEPS wave functions for the accessible values of $D$ include unbiased quantum fluctuations over short distance and high energy scales. Over longer length and lower energy scales, they revert to a more mean-field-like description. Thus, with the currently accessible $D$'s \cite{comment:D}, we cannot expect them to resolve between, e.g., static versus fluctuating stripes. On the other hand, our results provide significant evidence that approaches that ignore stripes do not give reliable descriptions of the $t$-$J$ model.



We acknowledge inspiring discussions with F. Becca, T. Nishino, and H.-Q. Zhou and funding from the Swiss National Science Foundation and from the NSF under DMR-0907500. The simulations were performed on the Brutus cluster at ETH Zurich.


\begin{thebibliography}{100}
\bibitem{commentSB} Since we consider systems of infinite size, symmetries may be spontaneously broken.
\bibitem{commentHuan}
Although charge- and spin-density waves have already been found in iPEPS simulations with $D=6$ \cite{Li10}, stripes could not be represented due to a too small unit cell.
\bibitem{commentMERA} Competitive variational energies (compared with VMC) have recently also been obtained with the 2D MERA for the Heisenberg model on the kagome lattice \cite{Evenbly09}.
\bibitem{comment:D} A larger bond dimension $D$ can be achieved by exploiting symmetries of the model\cite{symmetries} (we only used parity conservation in the present simulations). For a fixed $D$ the results can also be improved by going beyond the simple update used in the present work (cf. Ref.~\onlinecite{Corboz10b}).
\end{thebibliography}
\end{document}